\documentstyle [here,epsf,11pt,psfig] {article}
\begin {document}
\title    {On the Masses of the Physical Mesons: \break
           Solving the Effective QCD--Hamiltonian by DLCQ}
\author{Hans-Christian Pauli and J\"org Merkel \\ 
        Max-Planck-Institut f\"ur Kernphysik \\
        D-69029 Heidelberg \\ }
\date{2 December 1995}
\maketitle

\begin{abstract}
The effective QCD-interaction as obtained from the front form
Hamiltonian by DLCQ is Fourier transformed on the Retarded
Schr\"odinger Equation for to describe the constituents 
of physical mesons.
The crucial point is the use of a running 
coupling constant $\alpha_s(Q^2)$, in a manner similar but not equal 
to the one of Richardson in the equal usual-time quantization. 
Fixing the running coupling constant at the Z mass, 
the only parameters are the flavor masses. 
Without the top quark one needs thus 5 parameters to calculate 
the physical masses of 30 pseudoscalar and vector mesons, 
consistently within the same model. 
Applying variational methods to a caricature of the model, 
the biggest technical challenge is 
the solution of a cubic algebraic equation.~-- %
In view of an oversimplified model and a very simple 
technology, the agreement with the empirical data is much too good.
\end{abstract}

\vfill 
\noindent  Preprint MPIH-V45-1995 

\newpage 
\tableofcontents 

\newpage 
\section{Introduction and Motivation}

One of the most outstanding tasks in strong interaction physics is to 
calculate the spectrum and the wavefunctions of physical hadrons 
from Quantum Chromodynamics.
The method of `Discretized Light-Cone Quantization' (DLCQ) \cite{PaB85} 
has precisely this goal and has three major aspects: 
\begin {list} { } {\topsep 0ex\itemsep 0ex\parsep 0ex\leftmargin 2cm}
\item[(1)] a rejuvenation of the Hamiltonian approach, 
\item[(2)] a denumerable Hilbert space of plane waves, 
\item[(3)] Dirac's front form of Hamiltonian Dynamics.
\end {list} 
In the {\em front form}, 
using Dirac's terminology \cite{Dir49}, 
one quantizes at equal `light-cone time' $x^+=x^0+x^3$, 
as opposed to the {\em instant form} 
where one quantizes at equal `usual time' $t= x^0$. 
The front form carries many names like for example 
{\em light-cone quantization} \cite{Wei66}, see also \cite{BrP91}.
The approach has some unique features as reviewed in \cite{BrP91}:
The vacuum is simple, or at least simplier than in the instant form,
which in turn implies that the groundstate of the free theory 
is also the groundstate of the interacting theory. 
The relativistic wavefunctions transform trivially under certain 
boost operations \cite{Dir49,BrP91}.
Both are in stark contrast to the conventional instant form.
Discretization, that is the use of periodic or 
antiperiodic boundary conditions, provides additional length 
scales as useful {\em regularization parameters}
in a formalism like a quantum field theory 
which otherwise is littered with divergencies {\it ab ovo}. 
In fact discretization (or working `on a circle') is more 
than only a useful tool for calculations, rather it provides 
a systematic scheme to formulate the theory {\em at all}.

Over the years, the light-cone community \cite{Gla95} has made 
much progress. Calculations have been done \cite{BrP91} to verify
agreement with other methods particularly lattice gauge theory.
We have learned that {\em zero modes} should not be 
ignored \cite{HKS92}, since they can be important carriers 
of quantum structure. 
Other calculations in 1-space and 1-time dimension like the 
dimensionally reduced models \cite{DKB94,PKP95} 
have an interest of their own.
We have learned that there are physical and non-physical gauges, 
even something like a wavefunction of the vacuum
\cite{PKP95}. 
Renormalization is a difficult problem within
a Hamiltonian theory and worth the effort of the 
best \cite{PHW90,WWH94}. 

What was {\em not} shown thus far, however, 
is why all that is useful.
Particularly one lacks the contact to phenomenology beyond
the perturbative regime \cite{BrP91}.
We believe that one needs more work like 
\cite{BSW87} or like \cite{BrS94,Sch94} which relate 
formalism and actual experimental numbers.
It should be possible to combine rigor with simplicity!
The present work will hopefully be a useful contribution
to the extent that we shall derive a rather simple, conventional
second order differential equation for the wavefunction 
in configuration space. 
Amazingly enough, it seems possible to describe quarks in hadrons 
by a non-relativistic Schr\"odinger equation to account  
for the correct retardation 
and still being in the front form of 
Hamiltonian Dynamics applied to Quantum Chromodynamics.

Solving this `Retarded Schr\"odinger Equation', 
or better a caricature of this equation, 
by parametric variation, we end up with the masses 
of all pseudoscalar and vector mesons. 
In comparison with the experimental data \cite {PDG94}, 
they are no worse 
than conventional phenomenological models \cite {QuR79,GoI85}, 
or predictions from heavy quark symmetry \cite {Neu93}, 
or even predictions based on lattice gauge calculations
\cite {Mac93,BCS94,Wei94}. 
In view of the simplicity of the methods applied 
we regard this as a success of the front-form approach, 
particularly since we can and will do better in the 
future \cite{MeP96}.

\section{The Model: The Effective QCD-Hamiltonian from DLCQ} 

In Discretized Light-Cone Quantization one addresses oneself 
to solve an eigenvalue problem 
\begin{equation}
   H _{\rm LC} |\psi_i\rangle = E _i |\psi_i\rangle
\ . \label {eq:1} \end{equation}
The operator $ H_{\rm LC} \equiv  P^{\mu}P_{\mu} $ is a Lorentz-scalar
with the dimension of a $ <\!\!mass\!\!>^2$. 
The eigenvalues have the same dimensions. 
We therefore set $E _i = M _i ^2$ and interprete $ M _i $ as the 
`mass of the state $i$'. 
In DLCQ one works in momentum representation, and thus three of the
four components of $ P ^\mu$ are diagonal, namely the three total 
(light-cone) momentum operators $ P ^+ $ and $ \vec P _{\!\bot} $.
The fourth component, the total (light-cone) energy operator $ P ^-$
is very complicated and off-diagonal. 
The four components of energy-momentum $ P ^\mu $ mutually commute; 
their matrix elements can be found explicitly in \cite{BrP91}. 
The eigenvalues of $ P ^+ $ and $ \vec P _{\!\bot} $ are 
\begin{equation}
         P^+ = \sum _j k ^+ _j 
    \ , \quad {\rm and} \quad
    \vec P _{\!\bot} = \sum _j \vec k _{\!\bot _j }  
\ , \label {eq:2} \end{equation}
where $j$ runs over all partons (particles) in a Fock state;
each parton has a four-momentum denoted by 
$ k ^\mu _j = ( k ^- _j , k ^+ _j , \vec k _{\!\bot _j }) $ 
and sits on its mass shell $ ( k ^\mu k _\mu) _j = m _j ^2$.
Based on the boost-properties of light-cone operators \cite{BrP91}
one always can transform to an intrinsic frame, where
$ \vec P _{\!\bot} = 0$ and thus $ P ^\mu P_\mu = P ^+ P ^- $.
Since $ P ^+$ is diagonal from the outset, the diagonalization
of $P ^-$ and of $ H _{\rm LC} $ amounts to the same problem.

\begin{figure}[H]
\centerline{
\psfig{figure=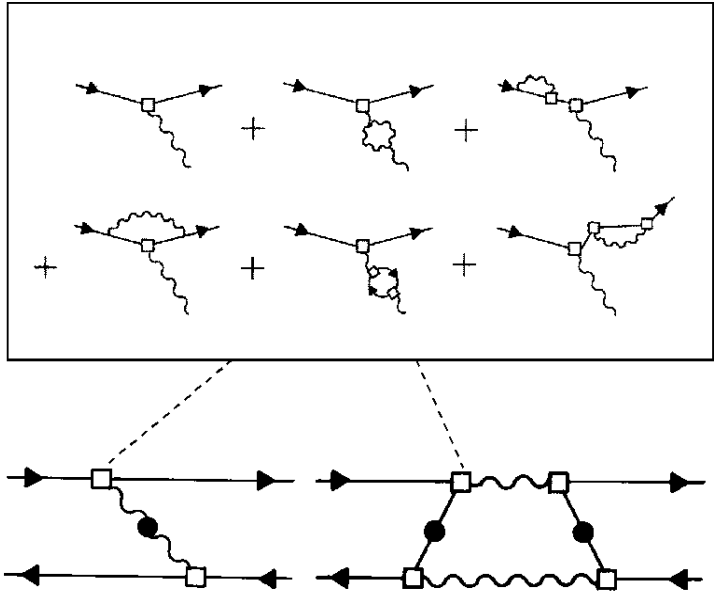,height=8cm,rheight=8cm,angle=0}
}
\vspace*{4cm}
\centerline{ \parbox{35em}{
   \caption{\label{fig:1} 
   \sl The effective interaction in the $q\bar q$ sector.~--- %
   \rm By absorbing (or emitting)  an `effective gluon', 
   a single quark state with four-momentum $ k _1 $ and 
   spin-projection $ s _1 $ is scattered into the quark state 
   ($ k _1 ^\prime , s _1 ^\prime $).
   Correspondingly, the antiquark is scattered from
   ($ k _2 , s _2 $) to ($ k _2 ^\prime , s _2 ^\prime $).
   } } }
\end{figure}

The Hilbert space for diagonalizing $P ^-$ is spanned by all 
possible Fock states which have fixed and given eigenvalues of 
$ P ^+ $ and $ \vec P _{\!\bot} = (0, 0) $. 
They can be arranged into `sectors' $ n = 1, 2, \dots, N $, 
corresponding to their parton composition.
We convene to call the sector with one quark and one antiquark
as the sector with $n=1$. The sector with one quark ($q$), 
one antiquark ($\bar q$), and one gluon ($g$) 
shall be denoted by $n=2$, and so on. 
It is peculiar to DLCQ that the number of sectors is limited 
for any fixed value of $ P ^+ $, or of 
the `harmonic resolution' $ K = \pi P ^+ /L$ \cite{PaB85}, 
since the longitudinal momenta $ k ^+ _j$ are all positive and 
non-vanishing numbers.
Both K and N can thus be arbitrarily large but are strictly finite.
The eigenvalue equation, Eq.(\ref{eq:1}), can thus be written 
as a matrix equation for a {\sl finite} number of block-matrices 
$ \langle n | H | n ^\prime \rangle $, {\it i.e.}
\begin{equation}
 \left(\begin{array}{cccc} 
    \langle 1| H |1\rangle & \langle 1| H |2\rangle & \ldots 
                           & \langle 1| H |N\rangle \\
    \langle 2| H |1\rangle & \langle 2| H |2\rangle & \ldots 
                           & \langle 2| H |N\rangle \\
            \vdots         &                        & \ddots 
                           &         \vdots         \\
    \langle N| H |1\rangle & \langle N| H |2\rangle & \ldots 
                           & \langle N| H |N\rangle 
 \end{array}\right)
 \left(\begin{array}{c} 
    \langle 1|\psi\rangle \\ \langle 2|\psi\rangle \\ \vdots 
                          \\ \langle N|\psi\rangle 
 \end{array}\right) 
  = M ^2
 \left(\begin{array}{c} 
    \langle 1|\psi\rangle \\ \langle 2|\psi\rangle \\ \vdots 
                          \\ \langle N|\psi\rangle 
 \end{array}\right)
\ . \label {eq:3} \end{equation}
The projected wavefunctions are denoted 
by $\langle n|\psi \rangle $.
The probability amplitude for finding only a $q\bar q$-pair 
with no gluons, in a pion for example, 
is denoted by $\langle 1|\psi \rangle=\psi_{q\bar{q}/\pi}$, 
and correspondingly for the higher sectors.
%
%
The Hamiltonian matrix Eq.(\ref{eq:3}) has a finite number of sectors, 
but thus far each sector has an infinite dimension, 
since the transversal momenta of each parton can take any positive 
or negative value, including zero. 
The number of states is however finite if one admits only those 
transversal momenta whose absolute values are smaller than a 
`cut-off'. The regularization of Brodsky and Lepage, {\it i.e.}
\begin{equation}
   \sum _{j\in n} 
   \Big({ m ^2 +\vec k _{\!\bot}^{\,2}\over x }\Big) _j 
   \leq \Lambda ^2 _n
   \ , \qquad{\rm with}\quad x _j = { k ^+ _j \over P ^+}
\ , \label {eq:4} \end{equation}
does that in a Lorentz-invariant way. 
The cut-off mass $\Lambda _n$ can but need not 
depend on the sector number.
In the $q\bar q$ sector, we shall use below 
$ \Lambda ^2 _1 = \Lambda ^2 + ( m _1 + m _2 ) ^2 $ 
such that the `dynamic' cut-off $ \Lambda $ 
is a measure for the off-shell momenta. 
The Hamiltonian matrix is then strictly finite: A finite number of
sectors, and within each sector a finite number of Fock states.
Now, any $N \times N$ block matrix equation with the 
Hamiltonian $H$ can be transformed identically 
\cite {Pau93} into a $1\times 1$ block matrix equation 
with the effective Hamiltonian $H _{\rm eff}$, 
{\it i.e.}
\begin{equation}
   H _{\rm eff} |\phi \rangle = M ^2 |\phi \rangle
   \ , \qquad{\rm with}\quad
   |\phi \rangle = |1\rangle \langle 1| \psi \rangle 
\ . \label {eq:5} \end{equation}
As a consequence of working with the front form, the light-cone
Hamiltonian \cite {BrP91} and therefore also the 
effective Hamiltonian is additive in the `kinetic energy' $T$ 
and the `effective interaction' $\widetilde U$, {\it i.e.} 
\begin {equation}
  M ^2 \; | \phi \rangle = T \; | \phi \rangle 
              + \widetilde U \; | \phi \rangle 
\ . \label {eq:6} \end {equation}  
Note that the effective Hamiltonian acts only 
in the sector with $n=1$, {\it i.e.} in the $q\bar q$ sector.
The reduction from Eq.(\ref{eq:4}) to Eqs.(\ref{eq:5}) 
or (\ref{eq:6}) is straightforward but complicated, 
involving only well-defined matrix inversions and 
multiplications \cite{Pau93}.
A number of points should be emphasized: No unphysical 
Fock-space truncations are needed 
as they are advocated sometimes as Tamm-Dancoff 
truncations \cite {Tam45,Dan50}. 
No smallness assumptions on the coupling constant 
is ever to be made. 
The algorithm expresses the `higher Fock space projections' 
$\langle n|\psi \rangle $ for $n \geq 2$ as a 
functional of $\langle 1 |\psi \rangle $.
Finally, one should emphasize that the procedure \cite {Pau93} 
is {\em only well defined} 
if the Fock space is denumerable like in DLCQ.
As a result one ends up essentially with two contributions to the 
effective interaction, as displayed in Fig.~\ref{fig:1}.
A more thorough presentation is presently being prepared.

Every finite and well-defined Hamiltonian matrix 
has thus a unique and well-defined effective Hamiltonian matrix. 
In the former \cite{KPW92} and the present work one considers
only the effective interaction as displayed 
on the left in Fig.~\ref{fig:1} 
and disregards (not neglects!) the annihilation term.
Without further arguing about that we consider the 
so defined effective interaction as part of our model.

DLCQ was particularly useful to convert the matrix 
equation (\ref{eq:3}) to another matrix equation (\ref{eq:6}). 
Once this is done, one can leave the {\sl discretized} case by 
going over to the {\it continuum limit}. 
In the continuum limit the matrix equation (\ref{eq:6}) is 
converted into the integral equation \cite {KPW92} 
\begin{eqnarray}
    M ^2 \; \phi _{s_1, s_2} (x, \vec k _{\!\bot})
 =  \Big( {m _1 ^2 + \vec k _{\!\bot} ^{\;2} \over x} 
        + {m _2 ^2 + \vec k _{\!\bot} ^{\;2} \over 1-x} \Big)
    \; \phi_{s_1, s_2}(x, \vec k _{\!\bot}) 
\nonumber \\
 + \sum_{s_1', s_2'}\!
    \int\!\! dx ^\prime \, d^2 \vec k _{\!\bot} ^\prime 
    \ { 2 m _1 m _2 \ \Theta ' 
    \over \sqrt{x(1-x)\, x'(1-x')} } 
    \ \widetilde U _{ s_1 , s_2 } ^{ s_1', s_2' } 
    (x , \vec k _{\!\bot} ; x', \vec k _{\!\bot}' )
    \ \phi_{s_1', s_2'}(x', \vec k _{\!\bot}')
\ . \label {eq:7} \end{eqnarray}
The kernel is written conveniently in terms 
of the interaction and the fermion currents,
\begin {eqnarray} 
\widetilde U = \widetilde V \widetilde C 
     \ , \qquad {\rm with} \qquad \qquad 
     \widetilde V (x , \vec k _{\!\bot} ; x', \vec k _{\!\bot}' )
 &=&  - \ {1\over 2\pi^2} \ \frac {\beta( Q _a ^2)} { Q _a ^2} 
\ ,  \nonumber\\ 
     {\rm and }\quad
     \widetilde C  _{ s_1 , s_2 } ^{ s_1', s_2' } 
       (x , \vec k _{\!\bot} ; x', \vec k _{\!\bot}' )
 &=&   \bar u (k_1, s_1) \, \gamma ^\mu\, u (k'_1, s'_1) 
     \ \bar u (k_2, s_2) \, \gamma _\mu\, u (k'_2, s'_2) 
\ .  \label {eq:9} \end {eqnarray} 
The cut-off function 
$ \Theta ' \equiv \Theta (x', \vec k _{\!\bot}') $
should be a remainder, that the domain of integration is limited 
according to Eq.(\ref{eq:4}).
The four-momentum transfer $Q ^2 _a $ is defined as the average 
of the quark and the anti\-quark, {\it i.e.} 
\begin {equation}
   Q _a ^2 = - q _a ^2 = -\frac {1} {2} (k' _1 - k _1)^2 
       - \frac {1} {2} (k' _2 - k _2) ^2
\ . \label {eq:10} \end {equation}  
Finally, $\beta$ is the dimensionless coupling constant.
For QED it takes the value $\beta = \alpha$  with $\alpha \sim 1/137$.
For QCD with 3 colors it takes the value $\beta = \frac{4}{3}\alpha_s$. 
In either case, it depends on the momentum transfer due to the
renormalization group. 

The kernel of Eq.(\ref{eq:9}) looks different from \cite {KPW92} 
without being so. 
Here, we shall work in the Bj\o rken-Drell
convention for the spinors such that in lowest order of 
approximation holds $ j ^\mu j _\mu \sim 1 $, 
as opposed to the Lepage-Brodsky convention \cite {BrP91} 
were one has $ j ^\mu j _\mu \sim 4 m _1 m _2 $.
In addition, a factor of $4 \sqrt{ x(1-x) x'(1-x') }$ is missing 
in the denominator due to a typing-error in \cite{KPW92}.~--  %
As illustrated in Fig.\ref{fig:1}, the effective 
interaction scatters a single quark with four-momentum $ k _1 $ 
and spin-projection $ s _1 $ into a quark with 
($ k _1 ^\prime , s _1 ^\prime $), and correspondingly an antiquark
from ($ k _2 , s _2 $) to ($ k _2 ^\prime , s _2 ^\prime $).
The eigenfunctions generated by Eq.(\ref{eq:7}) are identical with 
the probability amplitudes of the Fock-state expansion
for 3 colors, {\it i.e.}
\begin {eqnarray} 
     \vert \psi \rangle 
 &=& \sum_{s_1, s_2}
     \! \int\!\! dx \, d^2 \vec k _{\!\bot} 
     \, \phi _{s_1, s_2} (x, \vec k _{\!\bot})
     \, \vert x, \vec k _{\!\bot} ; s_1, s_2 \rangle 
\nonumber \\
     {\rm with} \quad 
     \, \vert x, \vec k _{\!\bot} ; s_1, s_2 \rangle 
 &=& {1\over\sqrt 3 } \sum _{c=1} ^3
        b ^\dagger _{c , s_1} (  x,   \vec k _{\!\bot}) 
     \, d ^\dagger _{c , s_2} (1-x, - \vec k _{\!\bot}) 
     \, \vert 0 \rangle 
\ . \label {eq:eif} \end {eqnarray} 
They depend on five entries: 
the three momenta of the quark and the two spin projections.
One may or may not consider Eq.(\ref{eq:7}) 
as the point of orign for the present model.
The principal aim of the present work is 
to generate solutions or reasonable approximations thereto. 

Having been so explicit, the spin projections and their summation 
will be suppressed in the sequel as being performed implicitly. 
The integral equation can then be written more compactly as
\begin {equation} 
    M ^2 \phi (x, \vec k _{\!\bot})
 =  \Big( {m _1 ^2 + \vec k _{\!\bot} ^{\;2} \over x} 
        + {m _2 ^2 + \vec k _{\!\bot} ^{\;2} \over 1-x} \Big)
    \phi (x, \vec k _{\!\bot}) 
 +  \int\!\! dx ^\prime \, d^2 \vec k _{\!\bot} ^\prime 
    { 2 m _1 m _2 \Theta ' 
    \ \widetilde U (x , \vec k _{\!\bot} ; x', \vec k _{\!\bot}' )
    \over \sqrt{x(1-x)\, x'(1-x')} } 
    \phi (x', \vec k _{\!\bot}')
\ . \label {eq:8} \end {equation}
In practice it is inconvenient to deal with the invariant 
mass-squared eigenvalues since binding effects are so much
overshadowed by the large value of the total free mass 
$ \overline M = m _1 + m _2 $. It is therefore suggestive to 
substitute the light-cone Hamiltonian $ H _{\rm LC} $ by 
an other operator $ H $, which differs by the former 
only by a constant and a readjustment of scale, {\it i.e.}
\begin {equation}
     H _{\rm LC} 
   = ( m _1 + m _2 ) ^2 + 2 ( m _1 + m _2 )\ H
\ . \label {eq:29} \end {equation}
The `Hamiltonian' $ H $ will then have the dimension of a
$ \langle mass \rangle $ or $ \langle energy \rangle $;
in fact, as we shall see, it will have much in common with
a non-relativistic Hamiltonian without being one.
Since $ ( m _1 + m _2 ) ^2 $ is a Lorentz scalar, the approach
continues to be frame independent. All what one does is to 
remodel the kinetic energy by 
\begin {equation} 
     T \equiv T (x , \vec k _{\!\bot} ) 
   = {1 \over 2 ( m _1 + m _2 ) } \left( 
           { m _1 ^2 + \vec k _{\!\bot} ^{\;2} \over x} 
         + { m _2 ^2 + \vec k _{\!\bot} ^{\;2} \over 1-x} 
         - ( m _1 + m _2 ) ^2 \right)
\ . \label {eq:31} \end {equation}
Without changing the substance, the integral equation becomes then 
\begin {equation} 
    E \; \phi (x , \vec k _{\!\bot} )
  = T (x , \vec k _{\!\bot} ) 
    \; \phi (x , \vec k _{\!\bot} ) 
 +  \int \! d^2 \vec k _{\!\bot} ^\prime 
    { m _r \ dx ^\prime \over \sqrt{x(1-x)\, x'(1-x')} } 
    \ \Theta ' \, \widetilde U 
           (x , \vec k _{\!\bot} ; x', \vec k _{\!\bot}' )
    \ \phi (x', \vec k _{\!\bot}')
\ . \label {eq:30} \end {equation}
The reduced mass $ m _r $ is given by
$ m _r = m _1 m _2 / ( m _1 + m _2 )$, as usually. 
The eigenvalue of invariant mass-squared $ M ^2 $ and the 
eigenvalue of the energy $ E $ are then related by
$ M ^2 = \overline M ^{\,2} + 2 \overline M E \not =  (\overline M + E ) ^2 $,
distinctly different from non-relativistic kinematics.

The running coupling constant does not depend on whether one 
calculates it perturbatively in the usual coordinates or 
by light-cone methods and behaves for sufficiently large 
momentum transfer like \cite {Pol73,GrW73} 
\begin{equation}
   \alpha_s( Q ^2) = \frac{12\pi}{33-2n_f}\, 
   \frac{1}{\ln( Q ^2/\kappa ^2 )}
    \ , \qquad\ {\rm for}\quad Q ^2 \gg \kappa ^2 
\ . \label {eq:11} \end{equation}
With the value of $\alpha _s$ as measured at the Z-mass \cite{PDG94} and 
three flavours ($n _f = 3$), one gets
\begin {equation}
  \alpha_s(M_Z)=0.1134 \pm 0.0035 
  \ , \qquad{\rm thus} \quad
  \kappa = 193\, {\rm MeV}
\ . \label {eq:12} \end {equation}  
The value of the `QCD-scale' $\kappa$ is thus fixed and will not be 
changed during this work.
But here is the problem: In a Hamiltonian approach like the present one, 
one needs the interaction not only at large but also at small momentum 
transfer, including zero. Taking the above functional at face value 
implies that hadronic interactions are {\it repulsive} 
for sufficiently small momentum transfer. 
This must plainly be wrong, of course, and the running 
must be modified.~-- But how?~-- 
Neither canonical renormalization theory nor the present 
DLCQ-formalism has a ready-to-use answer, yet.
We therefore modify it in the simplest possible way, 
inserting a constant at the right place: 
\begin{equation}
     \alpha _s ( Q ^2) = \frac{12\pi}{27}\, 
     \frac{1}{\ln(a ^2+ Q ^2/\kappa ^2 )}
     \ \qquad \qquad a \simeq 1
\ .  \label {eq:13} \end{equation}
Right or wrong, it does what one wants: The asymptotic behaviour is 
untouched and $ \alpha _s $ keeps its sign all the way down to 
$ Q ^2 = 0$. 
A similar step was taken by Richardson \cite{Ric79} for the instant 
form, but the value of $\kappa$ was considered a free parameter.
In the sequel we shall follow Richardson setting $a = 1$, 
we shall however go much beyond that 
by including the correct fine- and hyperfine interaction. 

The change from Eq.(\ref{eq:11}) to Eq.(\ref{eq:13}) 
is the only assumption in the present work: 
Despite not completely true, one could say that 
`confinement is introduced by hand' like in conventional potential
models. The crucial difference to other models is that the correct 
asymptotic freedom \cite{Pol73, GrW73} is combined with the virtue 
of having no free parameter and some further pleasant aspects: 
\begin {itemize}
\item
The model depends only on the {\it quark currents} and is therefore 
manifestly gauge-invariant. 
\item
The model is formulated in the front form and therefore 
frame- and boost-invariant. 
\item
The usual `recoil problem' is absent in a momentum representation.
\end {itemize}
Finally, one should mention here already that the Light-Cone formalism 
has a different operator structure than the conventional instant form: 
The total angular momentum $J$ and its eigenvalues cannot be used to 
classify the eigenstates \cite{{BrP91}}. 
The non-initiated reader might find it quite surprizing 
that the wavefunctions {\it can never be invariant} under
spatial rotations about the y- or the x-axis. 

\section{Analyzing the Model by Transforming Integration Variables}

Like in other models, the flavor quark masses are free parameters 
subject to be determined by comparing with experiment.
Natural candidates are the pseudoscalar ($0 ^-$) and vector 
mesons ($ 1 ^- $), which are believed to have total angular momentum $L=0$. 
Their total spin is thus $J=S$. 
For two quarks, the total quark spin $S$ 
can only take the value $S=0$ or $S=1$. 
Since the flavor quark masses potentially range from a few MeV up to 
some 100 GeV, one faces a tantalizing numerical problem: For any given set 
of quark and antiquark masses, $m _1$ and $m _2$, respectively, one
has to solve Eq.(\ref{eq:7}) and to perform a calculation of 
similar complexity as in \cite {{KPW92}}.
Actually, in addition to that, one is confronted with thus far 
unresolved, technically highly non-trivial problems due to a 
$ Q ^4$ singularity in the denominator. 
Moreover, of all the calculated {\it spectrum} one only needs 
the {\it lowest} eigenvalue.
Only the latter will be compared with the physical meson mass $M$.
Most of the calculated and thus expensive information is thus redundant,
not to mention that the calculation will have to be repeated a zillion 
of times in order to fit $m _1$ and $m _2$ to $M$.~--- 
Is there a simplier way? The following analysis will pave the way 
for an ultimate, possibly analytical solution.

The fundamental equation is unpleasant to the practitioner. 
In Eq.(\ref{eq:7}) two of the three integration variables are
dimensionful ($\vec k _{\!\bot} $) and run from $-\infty$ to $+\infty$, 
while the third one is dimensionless and has support only on the 
finite domain $0\leq x \leq 1$. 
It is therefore more than convenient to change variables from 
$x$ to $k_z$, such that $k_z$ has the same support and dimension as 
$\vec k _{\!\bot}$. This can be achieved by
\begin {equation} 
     x = x ( k _z) = { k _z + E _1 \over E _1 + E _2}
     \ , \qquad{\rm with}\quad 
     E _i = \sqrt {m _i ^2 + \vec k _{\!\bot} ^{\, 2} + \vec k _z ^{\, 2} }
          = \sqrt {m _i ^2 + \vec k ^{\, 2} }
     \quad{\rm for} \quad i = 1,2
\ . \label {eq:14} \end {equation} 
The physical interpretation of $k_z$ is of secondary importance. 
One notes with pleasure that 
$ k _z = 0$ and $ \vec k _{\!\bot} = 0 $ produces 
the correct `static value'  $\bar x = m _1 / ( m _1 + m _2)$
and that one can formally define a three-vector 
$\vec k = ( k _z, \vec k _{\!\bot} )$. 
The practitioner therefore will conclude that $ k _z $ is 
`something like' the relative momentum in $ z $-direction. 
The advantages are obvious, particularly the notation 
is more compact than in the standard front form. 
For example, the free part of the invariant mass squared appearing 
in Eqs.(\ref{eq:4}) and (\ref{eq:7}) 
becomes by direct substitution simply 
\begin {equation} 
    {m _1 ^2 + \vec k _{\!\bot} ^{\;2} \over x} 
  + {m _2 ^2 + \vec k _{\!\bot} ^{\;2} \over 1-x} 
  = ( E _1 + E _2) ^2
\ . \label {eq:15} \end {equation} 
{\it It looks as if} one changes the single particle four-momentum 
state from the front form with 
$ k ^\mu _1 = ( k ^- _1 , k ^+ _1, \vec k _{\!\bot _1} )$ and
$ k ^\mu _2 = ( k ^- _2 , k ^+ _2, \vec k _{\!\bot _2} )$, 
to a state in the instant form with 
\begin {equation} 
     k ^\mu _1 
   = ( k ^0 _1 , \vec k _{\!\bot _1} , k ^3 _1 )
   = ( E _1 , \vec k )
\ , \qquad{\rm and} \quad
     k ^\mu _2 
   = ( k ^0 _2 , \vec k _{\!\bot _2} , k ^3 _2, )
   = ( E _2 , - \vec k )
\ . \label {eq:16} \end {equation} 
The transfer and the mean of three-momentum 
along the quark line is then given by
\begin {equation} 
      \vec q    = \vec k - \vec k '
      \qquad{\rm and} \quad
      \vec p = {1\over2} (\vec k + \vec k ' )
\ , \label {eq:17} \end {equation} 
respectively. Along the antiquark line, both have the opposite sign. 
But all this does {\it not imply} that one works in the instant form.
Indeed, one has explicit residues from the front form, as follows.
Because of boost-invariance, Eq.(\ref{eq:7}) is independent
of the total longitudinal momentum $ P ^+$. 
But $ P ^+$ still lurks in the background, 
since the Fock-states `on the left' 
($ \vert x , \vec k _{\!\bot} \rangle $) 
have the same $ P ^+ $ than those `on the right' 
($ \vert x' , \vec k _{\!\bot}' \rangle $).
One can express this fact in instant form variables as well, using 
$ P ^+ = P ^0 + P ^3$. Since $ P ^3 = k _{z,1} + k _{z,2} = 0 $, 
one is left with $ P ^0 = P {^0} ' $,
and therefore has for every matrix element: 
\begin {equation} 
   E _1 + E _2 = E _1 ' + E _2 ' 
\ , \qquad {\rm or} \qquad 
   \vec k ^2 = \vec k ^{\prime\, 2} 
\ .\label {eq:18} \end {equation} 
Because of the latter, one has also individually 
$ E _1 = E _1 '$ and $ E _2 = E _2 '$. 
Obviously, the interaction {\it does not change the size}
of $ \vec k $, it only changes its direction: 
The frame-independent front-form formalism is equivalent 
to the center-of-mass frame of the instant form. 
This, of course, is a source of great simplification.
For example, the momentum transfer and the mean momentum 
are always orthogonal, 
\begin {equation} 
     \vec q \cdot \vec p 
   = {1\over2} ( \vec k - \vec k ' ) 
               ( \vec k + \vec k ' )
   = {1\over2} ( \vec k ^{\,2} - \vec k ^{\prime\,2} ) 
   = 0 
\ , \label {eq:19} \end {equation} 
and four-momentum transfer is always identical with the 
three-momentum transfer. The latter is verified easily. 
One inserts the single four-momenta from Eq.(\ref{eq:15}) 
into the defining Eq.(\ref{eq:10}) and gets 
\begin {equation} 
   Q _a ^2 = {1\over 2} \left[ 
           ( \vec k ' _1 - \vec k _1 ) ^2 - ( E ' _1 - E _1) ^2
         + ( \vec k ' _2 - \vec k _2 ) ^2 - ( E ' _2 - E _2) ^2
                     \right] = \vec q ^{\;2} 
\ , \label {eq:20} \end {equation} 
This way, the interaction $\widetilde V$ depends {\it only} 
on the spatial momentum-transfer.

\begin{figure}[H]
\centerline{
\psfig{figure=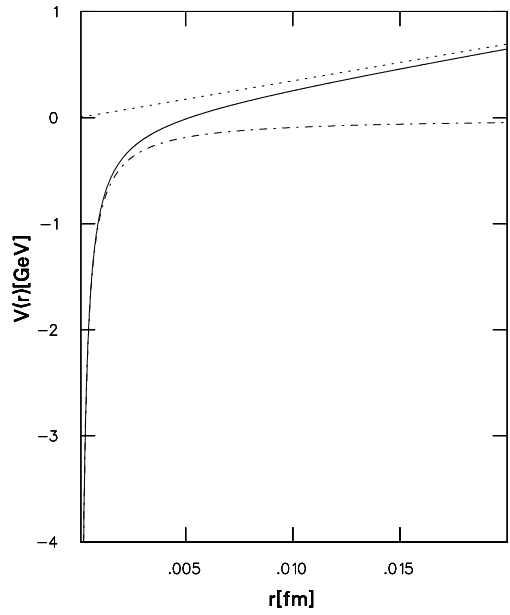,height=8cm,angle=0}
}
\centerline{ \parbox{35em}{
\caption{ \label {fig:2} 
   {\rm The quark-antiquark potential $V(r)$}. 
   {\sl Inserted is also the Coulomb (dashed-pointed) 
   and the confining potential (pointed).}
 } } }
\end{figure}

It therefore must be our aim to express all functions appearing
in the light-cone integral equation (\ref{eq:30}) 
as functions of the instant form variables $ \vec k $ and 
$ \vec k '$, {\it i.e.} to identically rewrite it as
\begin {equation} 
    E \,  \phi (\vec k \,) 
  = T (\vec k \,)\, \phi (\vec k \,) 
 +  \int \! d^3 \vec k ' 
    \  \Theta (\vec k ') 
    \, \widetilde A (\vec k , \vec k ' )
    \, \widetilde B (\vec k , \vec k ' )
    \, \widetilde C (\vec k , \vec k ' )
    \, \widetilde V (\vec k , \vec k ' )
    \, \phi (\vec k \,') 
\ . \label {eq:54} \end {equation}
The notation can be chosen such that the Jacobian 
$ \widetilde B $ and the current term $ \widetilde C $ 
are dimensionless with value `1' by lowest order 
of approximation, as will be unrevealed gradually.
Moreover, rather than by $ \vec k $ and $ \vec k '$, 
we shall express them often by the above 
$ \vec q $ and $ \vec p $, 
since they have a very special meaning in integral equations
like Eq.(\ref{eq:54}), as follows. 
%
%
Rewrite the latter conveniently as
$ E \phi ( \vec k )= \int\!\! d^3 \vec k ^{\,\prime} 
    \,\widetilde F (\vec q , \vec p ) \ \phi ( \vec k \,')$.
Multiply the whole equation with 
$ \exp ( i\vec k \cdot \vec x ) $ 
and subsequently integrate over $ d^3 \vec k $.
Making use of the definition 
\begin {equation} 
    \psi( \vec x \, ) = \int\!\! d ^3 \vec k 
    \ {\rm e} ^{ i\vec k \cdot \vec x } \ \phi (\vec k \,' ) 
    \ ,\qquad {\rm and\ thus} \quad 
    F (\vec x , \vec p \, ) = \int\!\! d ^3 \vec q
    \ {\rm e} ^{ i\vec q \cdot \vec x } 
    \ \widetilde F ( \vec q , \vec p \,) 
\ , \label {eq:22} \end {equation}
one straightforwardly arrives at an eigenvalue equation 
of the {\it Schr\"odinger type}, {\it i.e.} 
\begin {equation}
    E \psi (\vec x ) = 
    F (\vec x , \vec p \,) \ \psi ( \vec x )
   \ , \qquad {\rm with} \quad \vec p \equiv -i \vec \nabla _x
\ . \label {eq:23} \end {equation}
Thus, the analogue of $ \vec q $ is $ \vec x $, the position 
of the particle in the center-of-mass frame, and 
$ \vec p $ is its conjugate momentum. 
Consider for example the interaction term in Eq.(\ref{eq:54}).
Since $ \widetilde B = \widetilde C =1 $
will hold to lowest order, see below, one gets
\begin {equation}
      \widetilde V (\vec q^{\, 2})  
  = - {1\over 2 \pi ^2} 
        {\alpha_s (\vec q ^{\, 2}) \over \vec q ^{\, 2} }
      \ , \qquad {\rm thus } \quad 
      V( \vec x \, ) 
  =   {8\pi\over27} \Big( - \frac{1}{r} + \kappa^2\, r \Big)
\ . \label {eq:32} \end {equation}
The Fourier transform generates a Coulomb potential plus 
a confining potential with linearly rising walls, 
to a high degree of accuracy \cite{Ric79}.
This potential is plotted in Fig.~\ref{fig:2} versus 
$ r = \vert \vec x \vert$. 

Let us begin the programme with the kinetic energy 
$ T (\vec k \,) 
  = \left( ( E _1 + E _2 ) ^2 - \overline{M} ^2 \right)
    / (2 \overline{M} ) $. 
It is trivially a function only of $ \vec k $.
The cut-off function
$ \Theta ' = \Theta (\vec k \,') $ is related to
the mass-scale $\Lambda$ introduced in Eq.(\ref{eq:4}) 
to render the Hamiltonian matrix finite. The latter becomes now
$ ( E _1 + E _2 ) ^2 \leq \Lambda ^2 + ( m _1 + m _2 ) ^2 $,
which in turn can be cast identically into
\begin {equation}
     \vec k ^{\;2} \leq \left({\Lambda \over2}\right) ^2
     {\Lambda ^2 + 4 m _1 m _2 \over\Lambda ^2 + ( m _1 + m _2) ^2}
\ . \label {eq:25} \end {equation}
As expected, the cut-off is invariant under spatial rotations.
Next, using the Gordon decomposition, the currents 
$ \bar u (k_i, s_i)\, \gamma ^\mu \, u (k'_i, s'_i) $
are conveniently decomposed into their convective and spin parts, 
\begin {eqnarray} 
       \bar u (k_1, s_1)\, \gamma ^0 \, u (k'_1, s'_1) 
  &=&  { E _1 + m _1 \over 2 m _1 } \left(  1 
     + { \vec p ^{\,2} - \vec q ^{\,2}/4\over( E _1+ m _1) ^2} 
     + { \vec R \cdot\vec\sigma 
         \over( E _1 + m _1) ^2}\right) _{ s_1 , s_1'}
\ , \label {eq:62} \\
     \bar u (k_1, s_1)\, \vec \gamma \, u (k'_1, s'_1) 
 &=& { 1\over 2 m _1 } \left( 2 \vec p 
     - i\, \vec q \times\vec\sigma \right) _{s_1,s_1'}
     \ ,\qquad {\rm with} \quad
     \vec R = i \vec q \times\vec p  
\ . \label {eq:63} \end {eqnarray} 
Due to the validity of Eq.(\ref{eq:18})
these expressions are simpler than usually. 
Along the antiquark line one must change the sign of both $\vec k$ 
and $\vec k'$ and can conveniently substitute $ \sigma $ by $ \tau $ 
in order not to confuse the spin of quark and antiquark.
The current term $ \widetilde C = j ^\mu j_\mu $
becomes then explicitly
\begin {eqnarray} 
      \widetilde C ( \vec k , \vec k ' \,) 
  &=&\widetilde D \left(  1 
       + {\vec p ^{\,2}-\vec q ^{\,2}/4\over( E _1+ m _1) ^2} 
       + {\vec R \cdot \vec \sigma \over ( E _2 + m _2) ^2 } 
        \right) 
        \left(  1 
       + {\vec p ^{\,2}-\vec q ^{\,2}/4\over( E _1+ m _1) ^2} 
       + {\vec R \cdot \vec \tau \over ( E _2 + m _2 ) ^2 } 
         \right) 
\nonumber \\
  &+&\widetilde D \left( {2\vec p -i\,\vec q \times\vec\sigma
                \over E _1 + m _1 } \right) 
        \left( {2\vec p -i\,\vec q \times\vec\tau  
                \over E _2 + m _2 } \right) 
        \ ,\quad {\rm with} \ %
       \widetilde D = {( E _1+ m _1)( E _2+ m _2)\over 4 m _1 m _2 } 
\ . \label {eq:64} \end {eqnarray} 
As one expects for a Lorentz scalar it is invariant under spatial 
rotations.  
Finally, the Jacobian of the transformation Eq.(\ref{eq:14}) is 
evaluated in terms of the dimensionless transformation function 
$ \widetilde B ( \vec k , \vec k ' \,) $, implicitly defined by
\begin {equation}
      { m _r \over \sqrt{x(1-x)\, x'(1-x')} } \ dx ^\prime 
    = \widetilde A ( \vec k , \vec k ' \,) 
      \widetilde B ( \vec k , \vec k ' \,) \ dk _z
\ . \label {eq:52a} \end {equation}
By means of the identities 
\begin {equation}
      \frac {\partial x} {\partial k _z}
    = { ( E _1 + k _z ) ( E _2 - k _z ) \over 
           E _1 E _2 \, ( E _1 + E _2 ) }
\ \qquad {\rm and } \qquad 
    x (1 - x ) = { ( E _1 + k _z ) ( E _2 - k _z ) \over 
                                   ( E _1 + E _2 ) ^2 }
\ , \label {eq:33} \end {equation}
one gets straightforwardly
\begin {equation}
      \widetilde A ( \vec k , \vec k \,') 
    = \sqrt { {( E _ 1 + k _z ' )( E _ 2 - k _z ' )
         \over ( E _ 1 + k _z   )( E _ 2 - k _z   ) } }
\ , \qquad \qquad {\rm and} \quad
      \widetilde B ( \vec k , \vec k \,') 
    = \left( { m _r \over E _1 } + { m _r \over E _2 } \right)
\ . \label {eq:52} \end {equation}
Here $ \widetilde A $ is the only term 
which explicitly `breaks' rotational invariance 
since only here appears $ k _z $ explicitly. 
This is no failure, but must be true for any front form Hamiltonian. 
As a consequence, the wavefunction of the singlet state, for example, 
cannot be strictly invariant under rotations in three-space. 
One even has numerical evidence for such a `violation' 
due to the pioneering work of Krautg\"artner \cite{{KPW92}}:
The singlet wavefunction displayed in his Fig.~11 show them clearly.
First taken as a numerical artifact, they are meanwhile confirmed 
by an independent and improved calculation \cite{TrP96}. 
But these `violations' occur at a momentum scale much larger than 
the Bohr momentum, in a region where the wavefunction is down 
typically to $ 10 ^{-3} - 10 ^{-4} $ of its peak value. 
They seem to be unimportant.

Having  defined all terms in Eq.(\ref{eq:30}), it should be emphasized 
strongly that its solutions {\sl must be identical} with those obtained 
from the original light-cone equation (\ref{eq:7}), of course after 
having substituted $ k _z$ back by $ x $ according to Eq.(\ref{eq:14}).
It is however much simpler to deal with from the point of view of 
interpretation, of approximation, and last not least from the 
calculational point of view particularly for $ m _1 \not = m _2 $.

\section {The Retarded Schr\"odinger Equation}

The instant form has the garstly property of having square-roots 
scattered all over the place. 
This is particularly unpleasant when one aims at Fourier transforms. 
Can one find a systematic approximation scheme to remove them?~--
Once more this is possible by means of the Brodsky-Lapage cut-off. 
Indeed, one always can choose $\Lambda$ such that 
\begin {equation}
     {\vec k ^{\;2} \over m _1 ^2 } \leq 1
\ ,  \qquad {\rm for} \quad m _1  \leq m _2 
\ .  \label {eq:402} \end {equation}
If this holds for the lighter particle it even more so holds 
for the heavier one. All square-roots can then be expanded safely,
{\it i.e.} 
\begin {equation}
    E _i \simeq 
    m _i + { \vec k ^{\, 2} \over 2 m _i } 
\ , \qquad \qquad{\rm if }\quad 
    {1\over8} \left({\vec k ^{\,2}\over m _i^2}\right)^2 \ll 1 
\ . \label {eq:403} \end {equation}
This looks like a non-relativistic approximation but the
opposite is true: 
In the worst case, our choice allows for the ultra-relativistic 
velocities of the lighter particle up to 
$\vert\vec k \,\vert\sim m _1 $ on the one hand, 
and on the other hand for a consistent and systematic expansion 
up to second order like 
\begin {eqnarray} 
       \widetilde A 
   &=& 1 + {a q _z \over 2 m _r} 
         + { a ^2 q _z ^2 \over 8 m ^2 _r} 
         + { q _z p _z \over 2 m ^2 _r} 
            \left( a ^2 + { 2 m _r \over \overline M} \right)
   \ ,\ \qquad \qquad \qquad {\rm with} \quad 
   a = {m _1 - m _2\over m _1 + m _2} 
\ , \label {eq:405} \\
       \widetilde B 
   &=& 1 - {\vec p ^{\,2} \over 2 m _q ^2}
         - {\vec q ^{\,2} \over 8 m _q ^2 }
   \ , \qquad \qquad \qquad {\rm with} \quad 
   {1\over m _q ^2} = {1\over \overline M }
           \left( {m _2\over m _1^2} + {m _1\over m _2^2} \right)
\ , \label {eq:406} \\
   \widetilde D 
     &=& 1 + {\vec p ^{\,2} \over  4 m _a ^2 }
         + {\vec q ^{\,2} \over 16 m _a ^2 }
         \ ,\qquad \qquad \qquad {\rm with} \quad 
         {1\over m _a ^2 } = {1\over m _1^2} + {1\over m _2^2} 
\ , \label {eq:407} \end {eqnarray} 
respectively. 
The expansion of the current term $\widetilde C $ of Eq.(\ref{eq:64})
and of the product of the current with the Jacobian 
run analogously and give 
\begin {eqnarray} 
      \widetilde C 
   &=& 1 + {\vec p ^{\,2}\over 2}
           \left( {1\over m _a^2} + {2\over m _1 m _2} \right)
         - {(\vec\sigma \times \vec q\,) \cdot 
            (\vec\tau   \times \vec q\,) \over 4 m _1 m _2 }
         + { \vec\sigma \cdot \vec R \over 4 m _1 ^2}
         + { \vec\tau   \cdot \vec R \over 4 m _2 ^2}
         - { \vec S   \cdot \vec R \over   m _1 m _2}
\ , \label {eq:408} \\ 
       \widetilde B \widetilde C 
   &=& 1 + {\vec p ^{\,2}\over 2 m _1 m _2}
         - {\vec q ^{\,2}\over 8 m _q ^2}
         - {(\vec\sigma \times \vec q\,) \cdot 
            (\vec\tau   \times \vec q\,) \over 4 m _1 m _2 }
         + { \vec\sigma \cdot \vec R \over 4 m _1 ^2}
         + { \vec\tau   \cdot \vec R \over 4 m _2 ^2}
         - { \vec S   \cdot \vec R \over   m _1 m _2}
\ , \label {eq:409} \end {eqnarray} 
respectively. For completeness we finally define $\vec S$, 
the kinetic energy $T$, and $\Theta '$:
\begin {equation}
     \vec S = {1\over2} (\vec\sigma + \vec\tau) 
     \ , \qquad \quad
     T =  {\vec k ^{\,2} \over 2 m _r}
     \ , \qquad {\rm and} \quad
          \Theta (\vec k \,') = 1
\ . \label {eq:410} \end {equation}
The last step needs a comment. After having expanded in the
manner of Eq.(\ref{eq:403}) one has to verify in principle, 
that the results depend on the cut-off $\Lambda$ at most weekly.
One can conjecture however that the wavefunction decays sufficiently 
fast as to act itself like a cut-off. One thus can drop $\Lambda$ 
at this stage. All one has to do at the end of a calculation 
is to check the condition set in Eq.(\ref{eq:403}), but this 
can and will be done by an expectation value. 

One now is prepared to Fourier transform the whole Eq.(\ref{eq:54}) 
into an eigenvalue equation of the Schr\"odinger type
\begin {equation}
    H \psi( \vec x \,) = E \psi( \vec x \,) 
\ . \label {eq:411} \end {equation}
When transforming a momentum-space function like 
$ \vec R \widetilde V 
  = i ( \vec q \times \vec p \,) \widetilde V ( \vec q ) $
one gets in a first step
$ ( \vec \nabla \times \vec p \,) V $.
By definition, the operator $\vec \nabla$ 
can act {\it only} on the next $ V $ to the right.
Since $V$ is spherically symmetric one sets 
$ \displaystyle \vec \nabla V = {\vec x \over r} 
  {\partial V \over \partial r} $ and absorbs $ \vec x $ into 
the angular momentum operator 
$ \vec L \, =  \vec x \times \vec p $.
The Hamiltonian operator $ H $ in `Schr\"odinger representation'
(with $ \vec p = -i \hbar \vec \nabla $) turns then out as
\begin {eqnarray} 
    H 
   &=&  { 1 \over 2 m _r } 
        \left (1 + { V (r)\over m _1 + m _2} \right )
        \vec p ^{\,2}  
     +  V (r) 
        + {\vec \nabla ^{2} V (r) \over 8 m _q ^2}
        +  {(\vec\sigma \times \vec \nabla) \cdot 
            (\vec\tau   \times \vec \nabla V ) 
             \over 4 m _1 m _2 }
\nonumber \\ 
       &+& {1\over r} {\partial V \over \partial r } \left (
           { \vec\sigma \cdot \vec L \over 4 m _1 ^2}
         + { \vec\tau   \cdot \vec L \over 4 m _2 ^2}
         - { \vec S   \cdot \vec L \over   m _1 m _2} \right )
         + H _a
\ . \label {eq:412} \end {eqnarray} 
For a spherical $ V $ holds 
$ (\vec\sigma \times \vec \nabla) \cdot 
            (\vec\tau \times \vec \nabla V ) 
  = {2\over3} (\vec\sigma\cdot\vec\tau) \vec \nabla ^{2} V$ 
and with $ \vec\sigma\cdot\vec\tau = 2 \vec S ^{\,2} - 3 $ 
one arrives at 
\begin {eqnarray} 
     H 
    &=& { 1 \over 2 m _r } 
        \left (1 + { V (r)\over m _1 + m _2} \right )
        \vec p ^{\,2}  
     +  V (r) 
        -              {\vec\nabla ^{2} V \over 8 m _r ^2}
           \left( {3m _r\over m _1 + m _2}
           - {( m _1 - m _2)^2\over( m _1 + m _2)^2} \right)
        +  {\vec\nabla ^{2} V \over 3 m _1 m _2 } \vec S ^{2} 
\nonumber \\ 
       &+& {1\over r} {\partial V \over \partial r } \left (
           { \vec\sigma \cdot \vec L \over 4 m _1 ^2}
         + { \vec\tau   \cdot \vec L \over 4 m _2 ^2}
         - { \vec S   \cdot \vec L \over   m _1 m _2} \right )
         + H _a
\ . \label {eq:414} \end {eqnarray} 
The rotationally asymmetric $ H _a $ is given below.
We emphasize here already that the precise form of $ V $ need 
not be known in performing the step from Eq.(\ref{eq:409}) to 
Eq.(\ref{eq:414}). If one sticks to QCD in the above form, 
Eq.(\ref{eq:32}), one has
\begin {equation}
     \vec \nabla ^{2} V (r) = \beta 
     \left ({2\kappa ^2 \over r } + 4\pi\delta(\vec x )\right )
\,   \qquad {\rm and} \quad
     {1\over r} {\partial V (r) \over \partial r } 
   = \beta \left ( {\kappa ^2 \over r} + {1 \over r ^3} \right )
\ ,  \qquad {\rm with} \quad 
     \beta = {8\pi\over 27} \simeq 0.93
\ .  \label {eq:416} \end {equation} 
If one sticks to QED, one chooses $\beta\simeq 1/137$ and 
sets the QCD-scale $\kappa$ to zero.
These equations are at the core of the present work. 
They are still identical with the original equations (\ref{eq:30})
or (\ref{eq:54}) 
and include the correct retardation without smallness asumptions
particularly on the coupling constant or the mass parameters.

The `Retarded Schr\"odinger Equation' has 
a wonderfully simple structure and can be interpreted with ease.
One notes that the `potential' $V$ plays a different role in 
the different terms of the equation. 
The first term is the kinetic energy. Like in non-relatistic 
quantum mechanics the reduced mass governs the scale. 
Here the potential plays the role of an `effective mass'. 
In the second term $V$ appears in its natural role as a potential
energy with an attractive Coulomb and a linear string potential. 
In the second last term, the analogue of the `fine interactions' 
of atomic physics, the potential plays the role of a coupling
constant for the spin-orbit interaction.
Last not least, in the third and the fourth term one finds the
analogues of the atomic `hyperfine interaction' with the 
difference that they become of leading importance in the 
hadronic case. They contribute in the third term to the
true potential energy and for example modify the `Coulomb'
part of $V$, while the coefficient of $\vec S ^2$ can be 
interpreted as a (color-) magnetic interaction of the two 
spins. 

One should note that some parts of the Retarded Schr\"odinger 
equation are a direct consequence of working in the front form. 
We have tried to account for that in the notation. 
In Eq.(\ref{eq:408}), for example,
one notes that the current term has no contribution 
in $\vec q ^{\,2}$, while Eq.(\ref{eq:409}) 
does contain a term like that. They are caused by the 
explicit appearance of the light-cone momentum fraction  $x$, 
{\it i.e.} are caused by the Jacobian $\widetilde B \widetilde A $. 
The asymmetry Hamiltonian $ H _a $ has the same orign,
and following Eq.(\ref{eq:405}) has the Fourier transform 
\begin {equation}
  H _a 
     = { a _z \over 2 i m _r } {\partial   V \over\partial z } 
     - { a ^2 \over 8 m ^2 _r} {\partial^2 V \over\partial z ^2} 
     + {\partial V \over \partial z} 
       \left( a ^2 + { 2 m _r \over \overline M} \right)
       { p _z \over 2 m ^2 _r } 
\ .  \label {eq:420} \end {equation}
It looks like a self-induced hadronic interaction of the
quadrupole type.

How serious is the Retarded Schr\"odinger Equation to be taken?~-- 
We have actually no argument what could have been done wrong in 
the above, certainly nothing on purpose or with an unjustified 
assumption. At the very bottom the omission of zero-modes and, 
perhaps more important, the omission of
the two-gluon effective annihilation graph might perhaps be 
influential, and has to be investigated in the future. 
But not even the `Richardson potential' can be crucial.
Small distances $\vec x \rightarrow 0$ correspond to large momenta 
in Fourier space, and at least there asymptotic freedom must be correct. 
In the opposite limit, at very large distances corresponding to 
small momenta in Fourier space, the detailed form of the running 
coupling constant might possibly matter.
But there, the bound-state wavefunction is small from the outset 
and possible effects are unlikely to become important.
Not even the consistent expansion to second order in the momenta 
is an assumption, since it can and must be checked {\it a posteriori} by
\begin {equation}
     {1\over8} 
     \left ({\langle\vec p ^{\,2}\rangle\over m _i ^2}\right) ^2
     \ll 1
\ .  \label {eq:417} \end {equation}
One thus has to live with these equations. Rather than to agonize
further on these almost philosophical matters, to which we shall come
back in the summary, we proceed now by asking 
for practical consequences.

\section{Solving the Model by Parametric Variation}

It will take some time and effort to work out all the many consequences of
the Retarded Schr\"odinger Equation (\ref{eq:414}), which as 
mentioned should hold both for QED and QCD.
In the sequel, we therefore shall restrict ourselves to calculate 
only the ground-state masses of the pseudoscalar and  vector mesons. 
If one leaves aside the recently discovered top-quark $t$ and restricts to 
5 flavors ($u,d;s,c;b$), one has thus 30 different physical 
mesons, since charge-conjugate hadrons have of course the same mass. 

One cannot calculate these masses, however, without knowing the mass 
parameters $ m _1$ and $ m _2$ of the quarks. These cannot be measured 
in a model-independent way, although some people pretend one can do so.
Here, we shall adopt the point of view that the quark 
masses have to be determined consistently within each model, 
for the better or the worse. One has thus 5 mass parameters 
to account for 30 physical masses. Which ones should be selected 
to fit?~--- There are 142~506 different possibilities to select 
5 members from a set of 30, and we have to make a choice: 
We choose the five pure $q\bar q$-pairs. 
Even that is not unique: Shall one take the 
pseudoscalar ($0^-$) or the vector mesons ($1^-$)?~-- 
We shall take both, and compare the results. 

But doing so, one runs into the problem of choosing  
the chiral composition of the physical hadrons. 
One would be rather surprized if the pions, for example, 
could be understood as eigenstates in a simple potential model 
like Retarded Schr\"odinger Equation (\ref{eq:414}).
Thus, in the aim to avoid hell's kitchen by choosing 
a certain flavor composition, we shall do even worse and 
replace the `pions', for example, by `quasi-pions' 
with the same physical mass: All our hadrons shall be pure 
$q\bar q$-pairs~-- {\it by fiat}. For example, we shall identify the 
$u\bar u$-, $u\bar d$-, $d\bar u$-, or $d\bar d$\--eigenstates 
with the quasi-$\pi^0$, quasi-$\pi^+$, quasi-$\pi^-$, or the
quasi-$\eta$, and so on. These `crimes' can be revoked easily 
in future though not the present work. They are by no means 
compulsory to the model, as we shall see.

Our problems lie in another ball park. 
It is clear that we cannot deal head-on with the full complexity of 
the Retarded Schr\"odinger Equation (\ref{eq:414}).
Which part of the Hamiltonian should one select in a first 
assault?~-- Some help is gained by the rather unique property of the light-cone
Hamiltonian: Kinetic energy and interaction are additive, 
and so is the Hamiltonian. 
Because of this additivity one always can select an `interesting
part' $H_0$, {\it i.e.} $ H  = H _0 + \Delta H $  
and check {\it a posteriori}, 
by calculating $\langle \, \Delta H \,\rangle$ in perturbation theory, 
whether the selection makes sense, or not. 

Since the scalar and pseudoscalar mesons have probably no
orbital excitations, {\it i.e.} are primarily $s$-waves, 
one can disregard the spin-orbit couplings and choose
\begin {equation} 
 H _0  = \left (1 + { V (r)\over m _1 + m _2} \right )
               { \vec p ^{\,2} \over 2 m _r } 
     +  V (r) 
        -              {\vec\nabla ^{2} V \over 8 m _r ^2}
           \left( {3m _r\over m _1 + m _2}
           - {( m _1 - m _2)^2\over( m _1 + m _2)^2} \right)
        +  {\vec\nabla ^{2} V \over 3 m _1 m _2 } \vec S ^{2} 
\ . \label {eq:501} \end {equation} 
Even that looks to complicated for a start-up. 
We therefore select those terms which in one way or the other
have turned out to be important in past phenomenological
work, namely a simple central potential plus the triplet-interaction
mediated by the total spin. Our first choice is therefore
\begin {equation} 
  H _0  = { \vec p ^{\,2} \over 2 m _r } + V (r) 
     + {2 \kappa \over 3 m _1 m _2 } {\vec S ^{2} \over r} 
     = { \vec p ^{\,2} \over 2 m _r } 
     - {\beta\over r }
     + {2 \kappa ^2\over 3 m _1 m _2 } {\vec S ^{2} \over r} 
     +  \beta \kappa ^2 r 
\ . \label {eq:502} \end {equation} 
It is always possible to choose a spinorial representation in which
$ S _z$ and $ \vec S ^2$ become diagonal. In this case,
the latter can be replaced by the eigenvalue $ S _e = S( S + 1)$, 
taking the values 0 and 2 for the singlet and the triplet, 
respectively. 

%
%
How does the wavefunction for the lowest state look like?~--
For a pure Coulomb potential, {\it i.e.} for $\kappa = 0$, 
the solution has the form
\begin {equation}
  \psi (\vec x ) 
    = {1\over \sqrt{\pi} } \lambda ^{{3\over2}} 
      \, {\rm e} ^{-\lambda\, r} 
\ . \label {eq:503} \end {equation}  
One the other hand, ommitting the Coulomb part, 
a linear potential be solved in terms of Airy functions or its 
integral transforms \cite{PaB79}. If one has both, 
one will have some mixture of the two. In a numerical calculation, 
these can be found with almost arbitrary accuracy. 
But for the present start-up check even that requires too much effort. 

\begin {table}[t]
\begin {center}
\caption [masses] {\label{tab:1} {\sl
   The flavor quark masses in MeV, 
   as obtained from a fit to Eq.(\ref{eq:508}).~--- The 
   first row refers to a fit for the singlets, 
   the second to the one for the triplet.}}
\vspace{1em}
\begin {tabular}  {|c||c|c|c|c|c|}
\hline 
                  {\rule[-3ex]{0ex}{7ex}}
                  {\rm Flavor\ mass} 
                  & u & d & s & c & b 
\\ \hline \hline   
                  {\rule[-2ex]{0ex}{4ex}}
                  {\rm From\ fit\ to\ $0^-$} 
               \  & 2.3   & 155.6 & 430.6 & 1642.3 & 5330.8 
 \\ 
                  {\rule[-2ex]{0ex}{5ex}}
                  {\rm From\ fit\ to\ $1^-$} 
               \  & 222.8 & 236.2 & 427.2 & 1701.3 & 5328.2 
 \\ \hline 
\end {tabular}
\end {center}
\end {table}
Rather we shall pursue a variational approach. We
choose Eq.(\ref{eq:503}) as a one-dimensional parameter family,
subject to determine the one free parameter $\lambda$.
Of course, one could take other families as well, like
for example harmonic oscillator eigenstates, but with the above
all expectation values are particularly simple, {\it i.e.} 
\begin {equation}
     \langle\psi\vert\, \vec p ^{\,2} \,\vert\psi\rangle 
    = \lambda ^2 
\ , \qquad 
     \langle\psi\vert\, {1\over r} \,\vert\psi\rangle 
    = \lambda 
\ , \qquad {\rm and} \quad
     \langle\psi\vert\, r \,\vert\psi\rangle 
    = {3\over 2 \lambda }
\ . \label {eq:504} \end {equation}  
That's all what one needs to calculate the expectation value
of the energy 
\begin {equation} 
   \overline E = \langle\psi\vert H _0 \vert\psi\rangle 
     = { \lambda ^2 \over 2 m _r } 
     - \beta\lambda 
     + {2 S _e \kappa ^2 \over 3 m _1 m _2 } \,\lambda 
     +  {3\beta \kappa ^2 \over 2 } {1\over\lambda} 
\ . \label {eq:505} \end {equation} 
Since we deal only with a variational approach to the ground states, 
we are not in conflict with the statement 
that the wavefunction cannot be purely Coulombic.
For the pure Coulomb case, the 2$S$- and 1$P$-states would be 
degenerate and the ratio 
$|\psi _{2S}(0)| ^2/|\psi _{1S}(0)| ^2 = 0.125$, 
as opposed to the values $\simeq 0.63$ and $\simeq 0.50$ 
extracted from experiments on 
charmonium and bottomium \cite{QuR79}, respectively.

Since we address ourselves to calculate the total invariant mass 
of the hadrons we return to the front form invariant mass-squared
operator $ H _{LC} = M ^2$, {\it i.e.} to 
$ M ^2 = ( m _1 + m _2) ^2 + 2 ( m _1 + m _2) \overline E $.
Specializing now to equal masses $ m _1 = m _2 = m $ one
preferably converts the variational equation (\ref{eq:505}) 
in terms of the dimensionless variables
\begin {equation} 
     s = \frac{\lambda}{\kappa} 
\ , \qquad
   \xi = \frac{m}{\kappa} 
\ , \qquad {\rm and} \quad
     W = \left( \frac { M } {2\kappa} \right) ^2 
\ . \label {eq:506} \end {equation}   
We measure thus all energies and masses in units of the one 
fixed scale in our problem, the QCD scale $\kappa$. 
The variational eigenvalue equation reduces then to the 
handy expression
\begin {equation}
  W (s;\xi) = s ^2 
        - \left (\beta\xi - {2 S _e \beta\over3\xi} \right ) s
        + \xi ^2 
        + {3\beta\xi\over 2} \,{1\over s} 
\ . \label {eq:507} \end {equation}
At fixed values of the parameters ($\xi,\beta, S _e$), 
the variable $\lambda$ and thus $s$ must be varied such that
the energy (or the mass) become stationary, {\it i.e.}
\begin {equation} 
      {\partial W (s;\xi) \over \partial s } 
       \bigg\vert _{ s = s ^\star (\xi)} = 0
      \ ,\qquad {\rm thus} \quad 
      \left( { M \over 2\kappa} \right) ^2 
  =   W \left ( s ^\star (\xi) \right) = W ^\star (\xi) 
\ . \label {eq:508} \end {equation}
This leads to a cubic equation in $ s $ which can be solved
analytically in terms of Cardan's formula. 
In special cases they can well be approximated by a quadratic equation, 
namely when $s^\star \gg 1$ or when $ s ^\star \ll 1$.
We got accustomed to refer to these {\it two regimes} 
as the {\sl Bohr } and the {\sl string regime}, respectively. 
In the Bohr regime the Coulomb potential dominates the 
solution and the linear string potential provides a correction.
In the string regime the linear string potential dominates, 
with the Coulomb potential giving a correction.
Solutions in the string regime, however, imply that 
the ratio $ \langle \vec p ^{\,2} \rangle / m^2 = (\lambda/m) ^2$
becomes so large that one is in conflict with the 
validity condition Eq.(\ref{eq:417}).

\begin {figure}[H] 
\vspace {1em}
\centerline{\psfig{figure=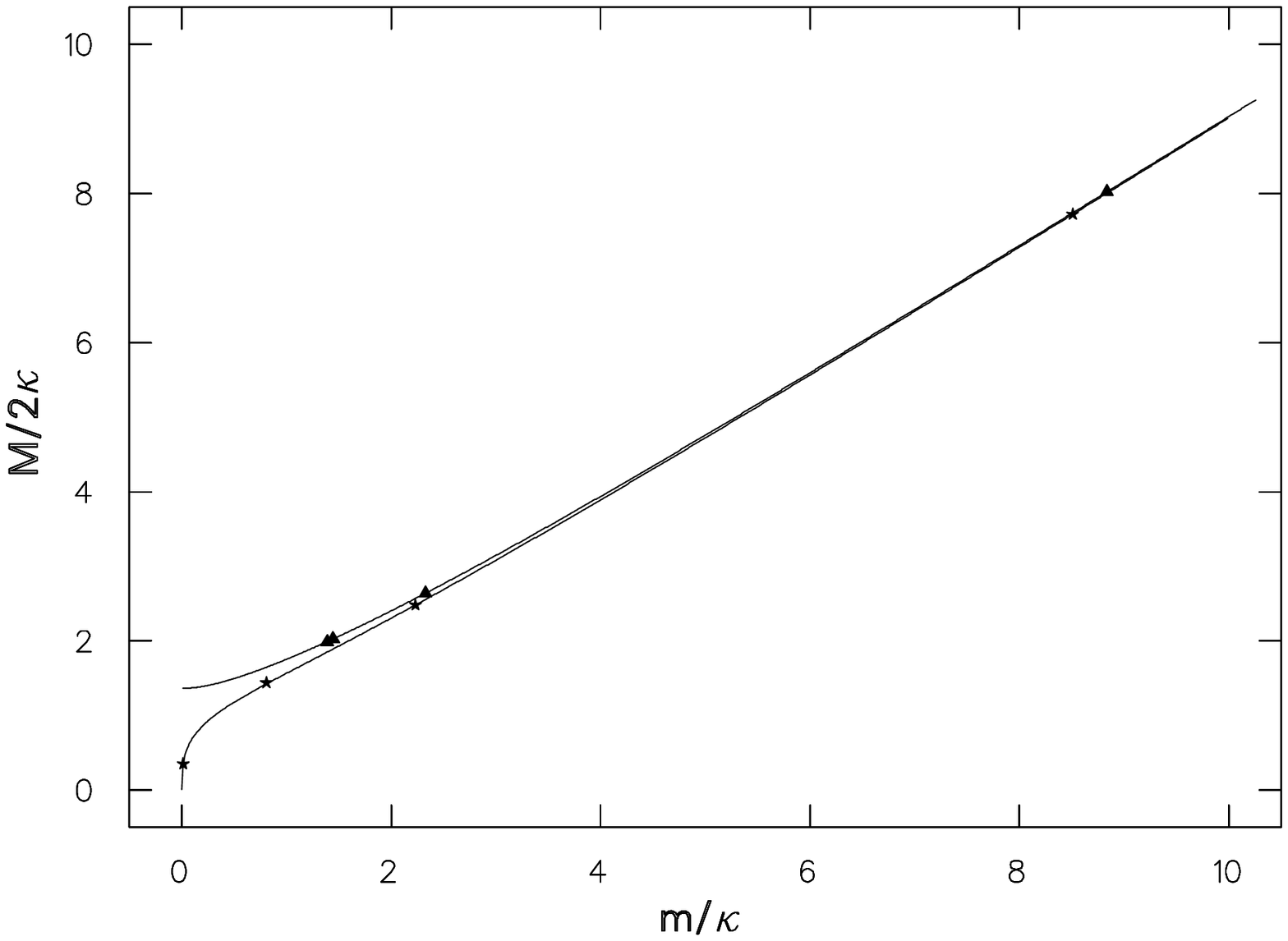,height=8cm,angle=90}}
\vspace {1em}
\caption {\label {fig:3} {\sl
    Bound states of $q\bar q$-pairs {\it versus} quark masses.~---  %
    All masses are given in units of the QCD-scale $\kappa$. 
    The upper curve refers to the triplet ($S_e=2$), 
    the lower to the singlet ($S_e=0$).
    The masses of some vectormesons ($\rho^0,\omega,\phi,J/\psi$) 
    are marked by ($\triangle$),
    those of the some pseudoscalars 
    ($\pi^0,\eta, \eta',\eta _c$) by ($\star$). }}
\end {figure}

\begin {table} [t]
\begin {center}
\caption [validity] {\label{tab:2} {\sl 
    The validity check.~--- 
    The first row displays the values of $(\lambda/m)^2 $ 
    as obtained from the mass fit to the singlets, 
    the second row those from the mass fit to the triplets.~---
    Note: If $(\lambda/m) ^4 \geq 8 $, the solution has to 
    be rejected as a consequence of Eq.(\ref{eq:417}), since
    $ \langle \vec p ^{\,2} \rangle = \lambda ^2 $. 
    The extremely large value for the $u$-quark in 
    pseudoscalar fit gives a good example for such a case. }}
\vspace {1em}
\begin {tabular}   {|c||c|c|c|c|c|}
\hline 
                  {\rule[-2ex]{0ex}{6ex}}
                  $\displaystyle {\lambda^2\over m^2}$ 
                  & u & d & s & c & b 
\\ \hline \hline   
                  {\rule[-2ex]{0ex}{5ex}}
                  {\rm For\ fit\ to\ $0^-$} 
                \ & 293  & 1.45 & 0.53 & 0.25 & 0.22 
 \\ 
                  {\rule[-3ex]{0ex}{6ex}}
                  {\rm For\ fit\ to\ $1^-$} 
                \ & 0.79 & 0.76 & 0.48 & 0.25 & 0.22 
 \\ \hline 
\end {tabular}
\end {center}
\end {table}
Rather than to give explicitly the straightforward but cumbersome 
formalism, we present the analytical results 
in the graphical form of Figure~\ref{fig:3},
both for the singlet and the triplet.
As displayed there, the total mass $ M = 2 \kappa \sqrt { W ^\star} $
is almost linear in the quark mass, with small but significant 
deviations. In line with expectation, 
the hyperfine splitting decreases with increasing quark mass.
Less expected was that the splitting increases so strongly 
with decreasing quark mass.
As one can see in the Figure, 
the mass of the singlet states takes off from the value 0
at vanishing mass almost like a square-root as opposed 
to the triplets which start off at a finite value.
Therefore, if one determines the mass of the $u$-quark by fitting 
to the quasi-pion mass, as displayed in Table~\ref{tab:1},
the value of $s$ becomes extremely small.
Consequently, one is in the string regime and the $u$-quark becomes 
ultra-relativistic. The validity condition is then badly violated, 
as compiled in Table~\ref{tab:2}. 
It is worth noting that the value of the up-quark mass is rather
close to what is reported for the so called `current mass'.
For the $\eta$ and the $\eta'$, the scaling variable $s$ is of 
order unity, while for the quasi-$\eta _c$ one definitely is in 
the Bohr regime.
Here the masses are similar or close to what is referred as 
the `constituent-quark' mass.

Since singlet and triplet are so close for $s\gg 1$ and since
nobody wants to fit the pion anyway, quasi or not, one fits
the quark masses preferentially with the vector mesons.
In the lack of empirical data we have set 
$ M _{\eta _b} = M _{\Upsilon} $, which should be of minor
importance in the present model, see 
Figure~\ref{fig:3}. 
As seen from Table~\ref{tab:1},
the flavor masses are now in close agreement with the 
constituent quark masses, particularly the up- and down-quarks.
As compiled in Table~\ref{tab:2}, the smallness condition
is now satisfied excellently, everywhere.

Obviously, it would be a minor modification  
to introduce some preferred chiral structure of the light mesons, 
and to drop the simple and simple-minded construction of quasi-mesons. 
But we have an other objection, here. We ask:
With the so fixed flavor quark masses, how bad or how good 
are the remaining 25 meson masses described?~--
The procedure runs quite analogously, except that it is now easier.
With fixed masses, one varies $\lambda$ in Eq.(\ref{eq:505}) 
such that the energy becomes stationary. Of course, this 
has to be done separately for each flavor composition.
We compile the results in Table~\ref{tab:4}. 
The table contains also a comparison with the experimental values,
to the extent the latter are known. 
It is clear why one should exempt the quasi-pions from a detailed
comparison with data, but for the rest it was not really anticipated
that they would agree on the level of about one part in hundred.
Some of these mesons have not yet been found.
The present model predicts then for example
\begin {equation}
    M (B _c ^{\pm})  = 6494 \,{\rm MeV}
    \ , \qquad 
    M (B _c ^{*\pm}) = 6501 \,{\rm MeV} 
\ . \label {eq:509} \end {equation}  
Such a table was not given before.
In all due respect for the work with lattice gauge theory,
the agreement of the above equations with data is not so much worse,
particularly in view of their simplicity and the absence of 
any free parameters.

\section {Summary and Conclusions}

Applying DLCQ to QCD in the light-cone gauge $ A ^+ = 0$ and 
disregarding {\em zero modes}, one derives first an effective 
interaction acting in the $q\bar q$ space. 
We emphasize that no assumption on a possible smallness of the 
coupling constant has to be made, 
nor that it is necessary to truncate the Fock space in the manner 
of Tamm \cite {Tam45} and of Dancoff \cite {Dan50}. 
Gauge invariance, therefore, is {\em not} violated. 
Once this is achieved, one can drop discretization and convert 
the matrix into an integral equation, Eq.(\ref{eq:8}). 
Unfortunately we cannot be more explicit in the present context.
We therefore consider this equation as part of the model and 
the point of departure in the present work.~-- %
We discover next a simple substitution of integration variables,
Eq.(\ref{eq:14}), which allows to transmute the equation identically 
into Eq.(\ref{eq:54}), in which appear only the conventional 
instant form variables like the usual momenta and energies.
One wishes to Fourier transform the still front form equations 
to usual space coordinates but is hampered by the `square roots'.
A poor man's solution is their consistent expansion 
up to second order. The prize to pay is Eq.(\ref{eq:417}), 
{\it i.e.} a smallness requirement {\it a posteriori} 
which has to be met in the solution. 
One ends up with a differential equation in configuration 
space, Eq.(\ref{eq:501}), which looks like a non-relativistic 
Schr\"odinger equation without being one. 
It contains the full retardation of a relativistically 
correct front form equation, and therefore is 
referred to as the Retarded Schr\"odinger Equation.
Its solutions can be Fourier transformed back to momentum space
and subjected to the transmutation equation (\ref{eq:14}), 
to get the wavefunction expressed in light-cone momenta 
particularly longitudinal momentum fractions, 
which are needed so urgently for the theoretical predictions 
of experimental high-energy cross sections, and 
of structure functions and the like.

One should emphasize that the Retarded Schr\"odinger 
Equation has no free parameter: 
The coupling constant and the quark masses have 
to be determined from the experiment.
Like in the familiar quantum electro-dynamics
one has no way to calculate them.
Fitting the strong coupling constant at the Z-mass 
and the 5 quark flavor masses to 5 selected vector 
mesons fixes all our freedom. 
The rest is structure: The 20 remaining 
pseudoscalar or vector masses are then predicted
and presented in Table~\ref{tab:4}.
In comparison with the experimental data they are 
not much worse than those 
from conventional phenomenological models \cite {QuR79,GoI85}, 
or from heavy quark symmetry \cite {Neu93}, 
or even from lattice gauge calculations \cite {Mac93,BCS94,Wei94}. 
The pions, of course, remain mysterious particles 
like in every other model not specially designed for them.
One wishes now to solve the full 
Retarded Schr\"odinger Equation (\ref{eq:501}) 
rather than only its caricature in Eq.(\ref{eq:502}).

Conclusion: If such a poor model can do so well
one must be on the right track. It seems that the front form
and DLCQ have made a big step forward. 
Unfortunately, it took ten years but 
one has been faced with all the problems of formulating 
a Hamiltonian approach to gauge field theory, 
in addition to the problems inherent to the front form.

\section{Acknowledgement}
HCP thanks Stanley~J.~Brodsky for the many discussions and 
exchange of ideas over all those ten years particularly 
for his patience in listening to the ideas still vague 
at the time of the Kyffh\"auser meeting.
``Of course'', he said, ``that's what Richardson did.''~-- In
the final phase of writing-up the content of the master 
thesis \cite{Mer94} we got to knowledge on similar ideas 
by Zhang \cite{Zha95}.

\newpage 

\newpage \setcounter{page}{ 25}
 \small\normalsize
\begin {table}[t]
\begin {center}
\caption [BigTableII] {\label{tab:4} \sl 
    The masses of $q\bar q$--hadrons are 
    compared with experimental values.~-- \rm
    Flavor quark masses used are inserted in column 2.
    They in turn come from a fit to the vector 
    mesons, as discussed in the text.~-- %
    Within each box, the first line refers to the 
    hadronic symbol of the meson; 
    the second line gives the calculated (measured) 
    vector mass in MeV;
    the third line accounts for the calculated (measured) 
    pseudoscalar mass in MeV, 
    and finally the fourth line specifies 
    the hadronic symbol of the pseudoscalar meson.~-- %
    Note: The smallness parameter $(\lambda/m)^2$ 
    satisfies always the condition set in the text. \rm }
\vspace{5mm}
\begin {tabular}  {|c|c||c|c|c|c|c|}
\hline
   {\rule[-3ex]{0ex}{7ex}}                   
   \qquad { }\qquad     & $\bf m_q$          & $\overline{\bf u}$ 
   & $\overline{\bf d}$ & $\overline{\bf s}$ & $\overline{\bf c}$ 
   & $\overline{\bf b}$ 
\\ \hline \hline
   & & $\rho^{0}$ & $\rho^{+}$ & $K^{*+}$ 
     & $\overline{D}^{*0}$ & $B^{*+}$ \\
   & & 768(768) & 773(768) & 910(892) & 2110(2007) & 5712(5325) \\    
   {\rule[-0.2ex]{0ex}{0.8ex}} 
   \bf u & 222.8 & & & & & \\ 
   & & 714(135) & & & & \\
   & & $\pi^{0}$ & & & & 
\\ \hline
   & & & $\omega$ & $K^{*0}$ & $D^{*-}$ & $B^{*0}$ \\
   & & & 782(782) & 914(896) & 2109(2010) & 5709(5325) \\
   {\rule[-0.2ex]{0ex}{0.8ex}} 
   \bf d & 236.2 & & & & & \\
   & & 658(140) & 668(549) & & & \\
   & & $\pi^{-}$ & $\eta$   & & & 
\\ \hline
   & & & & $\phi$ & $D_s^{*-}$ & $B^{*0}_s$ \\
   & & & & 1019(1019) & 2156(2110) & 5735(~---~) \\ 
   {\rule[-0.2ex]{0ex}{0.8ex}} 
   \bf s & 427.2 & & & & & \\ 
   & & 825(494) & 831(498) & 953(958) & & \\
   & & $K^{-}$ & $\overline{K}^{0}$ & $\eta^{'}$ & & 
\\ \hline
   & & & & & $J/\psi$ & $B^{*+}_c$ \\ 
   & & & & & 3097(3097) & 6502(~---~) \\
   {\rule[-0.2ex]{0ex}{0.8ex}} 
   \bf c & 1701.3 & & & & & \\
   & & 2079(1865) & 2078(1869) & 2131(1969) & 3082(2980) & \\
   & & $D^{0}$ & $D^{+}$ & $D_s^{+}$ & $\eta_c$ & 
\\ \hline
   & & & & & & $\Upsilon$ \\
   & & & & & & 9460(9460) \\
   {\rule[-0.2ex]{0ex}{0.8ex}} 
   \bf b & 5328.2 & & & & & \\
   & & 5701(5278) & 5698(5279) & 5726(5375) 
                  & 6495(~---~) & 9455(~---~) \\
   & & $B^{-}$ & $\overline{B}^{0}$ & $\overline{B}^0_s$ 
     & $B^-_c$ & $\eta_b$ 
\\ \hline   
\end {tabular}
\end {center}
\end {table}
 \small\normalsize
\newpage

\begin{thebibliography}{30}
\bibitem {PaB85} 
   H.C.~Pauli and S.J.~Brodsky, 
   Phys.~Rev. {\bf D32} (1985) 1993.
\bibitem {Dir49} 
   P.A.M.~Dirac, 
   Rev.~Mod.~Phys. {\bf 21} (1949) 392. 
\bibitem {Wei66} 
   S.~Weinberg, 
   Phys.~Rev. {\bf 150} (1966) 1313.
\bibitem {BrP91} 
   S.J.~Brodsky and H.C.~Pauli, in 
   \it Recent Aspects of Quantum Fields, \rm
   H.~Mitter and H.~Gausterer, Eds., 
   Lecture Notes in Physics, Vol 396, 
   (Springer, Heidelberg, 1991); and references therein.
\bibitem {Gla95} 
   \it Theory of Hadrons and Light-front QCD, \rm
   S.D.~Glazek, Ed., 
   (World Scientific Publishing Co., Singapore, 1995).
\bibitem {HKS92} 
   T.~Heinzl, S.~Krusche, S.~Simb\"urgen, and E.~Werner, 
   Z.~Phys. {\bf C56} (1992) 415. 
\bibitem {DKB94}
   K.~Demeterfi, I.R.~Klebanov, and G.~Bhanot, 
   Nucl.~Phys. {\bf B418} (1994) 15; 
   and references therein.
\bibitem {PKP95} 
   H.C.~Pauli, A.C.~Kalloniatis, and S.S.~Pinsky,
   Phys.~Rev. {\bf D52} (1995) 1176.
\bibitem {PHW90} 
   R.J.~Perry, A.~Harindranath and K.~Wilson, 
   Phys.~Rev.~Lett. {\bf 65} (1990) 2959.
\bibitem {WWH94} 
   K.G.~Wilson, T.~Walhout, A.~Harindranath, 
   W.M.~Zhang, R.J.~Perry, and S.D.~Glazek,
   Phys. Rev. {\bf D49} (1994) 6720;
   and references therein. 
\bibitem {BSW87} 
   M.~Bauer, B.~Stech, and M.~Wirbel,
   Z.~Physik {\bf D29} (1987) 103. 
\bibitem {BrS94} 
   S.J.~Brodsky and F.~Schlumpf,
   Phys.Lett. {\bf B329} (1994) 111.
\bibitem {Sch94} 
   F.~Schlumpf,
   J.~Phys. {\bf G20} (1994) 237;    and references therein.
\bibitem {PDG94} 
   L.~Montant, Ed., \it Review of Particle Properties, \rm
   Phys.~Rev. {\bf D50} (1994).
\bibitem {QuR79} 
   C.~Quigg and J.L.~Rosner, 
   Phys.~Rev. {\bf 56C} (1979) 167.
\bibitem {GoI85} 
   S.~Godfrey and N.~Isgur, 
   Phys.~Rev. {\bf D32} (1985) 189. 
\bibitem {Neu93} 
   M.~Neubert, 
   Phys.~Lett. {\bf C245} (1994) 259; 
   and references therein. 
\bibitem {Mac93} 
   P.B.~Mackenzie, 
   \it Status of Lattice QCD, \rm 
   In: Ithaka 1993, Proceedings, Lepton and Photon 
   Interactions, p.634; 
   \tt hep-ph 9311242; \rm 
   and references therein. 
\bibitem {BCS94}
   F.~Butler, H.~Chen, J.~Sexton, A.~Vaccarino, and D.~Weingarten,
   Nucl.~Phys. {\bf B430} (1994) 179. 
\bibitem {Wei94}
   D.~Weingarten,
   Nucl.~Phys. (Proc.~Supp.) {\bf B34} (1994) 29. 
\bibitem {Pau93} 
   H.C.~Pauli, in 
   \it Quantum Field Theoretical Aspects of High Energy Physics, \rm 
   B.~Geyer and E.M.~Ilgenfritz, Eds., 
   (Zentrum H\"ohere Studien, Leipzig, 1993);
   also available as preprint MPIH-V24-1993, Heidelberg, Oct 1993.
\bibitem {Tam45} 
   I.~Tamm, 
   J.~Phys. (USSR) {\bf 9} (1945) 449.
\bibitem {Dan50} 
   S.M.~Dancoff, 
   Phys. Rev. {\bf 78} (1950) 382.
\bibitem {Ric79} 
   J.L.~Richardson, 
   Phys.~Lett. {\bf 82B} (1979) 272.
\bibitem {KPW92} 
   M.~Krautg\"artner, H.C.~Pauli and F.~W\"olz, 
   Phys.~Rev. {\bf D45} (1992) 3755.
\bibitem {TrP96} 
   U.~Trittmann, and H.C.~Pauli, 
   ongoing work, to be published.
\bibitem {Pol73} 
   H.D.~Politzer, 
   Phys.~Rev.~Lett. {\bf 26} (1973) 1346.
\bibitem {GrW73} 
   D.~Gross and F.~Wilczek, 
   Phys.~Rev.~Lett. {\bf 26} (1973) 1343.
\bibitem {PaB79} 
   H.C.~Pauli and N.L.~Balazs,
   Math.~Comp.~{\bf 33} (1979) 353.
\bibitem {BjD66} 
   J.D.~Bj{\o}rken and S.D.~Drell, 
   \it Relativistic Quantum Mechanics, \rm 
   (McGraw-Hill, New York, 1964).
\bibitem {MeP96} 
   J.~Merkel and H.C.~Pauli,  
   \it work in preparation, \rm
   to be published in 1996.
\bibitem {Mer94} 
   J.~Merkel, 
   \it Diplomarbeit im Studiengang Physik, \rm
   U.~Heidelberg, Dec 1994.
\bibitem {Zha95} 
   W.-M.~Zhang, 
   Preprint IP-ASTP-19-95, Taipei, Oct 1995.
\end{thebibliography}
\end{document}